# Controlled Dephasing *via* Phase Detection of Electrons: Demonstration of Bohr's Complementarity Principle

Sprinzak, D.[1], Buks, E.[1,2], Heiblum, M.[1], & Shtrikman, H.[1]

[1] *Braun Center for Submicron Research, Department of Condensed Matter Physics, Weizmann Institute of Science, Rehovot 76100, Israel*

[2] *Condensed Matter Physics, California Institute of Technology, Pasadena, California 91125, USA*

**Interference results when a quantum particle is free to choose among a few *indistinguishable* paths. A canonical example of Bohr's complementarity principle [1] is a *two-path interferometer* with an external detector coupled to one of the paths. Then, interference between the two paths vanishes (*i.e. dephasing*) if one is able to detect, even in principle, the path taken by the particle. This type of *which path* (WP) experiment was already executed with photons, cooled atoms, neutrons, solitons, and more recently with electrons in a mesoscopic system [2]. In the latter experiment path determination was provided by inducing a change in the current flowing through the detector. In the present experiment we perform a more intriguing WP determination: only a phase change is being induced in the detector while the current remains unaffected. We show that such detector-interferometer interaction can be understood both, from the WP information obtained by the detector, or alternatively by understanding the disturbance in the interferometer caused by the detector. Moreover, we address the subtle role of detector coherency and point out difficulties in the intuitive interpretation of the experiment.**

An electronic mesoscopic system, where electrons maintain their wave properties, may serve as an excellent playground for demonstrating quantum mechanical interference. Moreover, Coulomb interaction among the electrons may facilitate strong *entanglement* (*i.e.*, quantum correlation) between coupled coherent systems – a useful quantum state in the emerging field of quantum computation. Recently, a mesoscopic *which path* (WP) type experiment was performed by Buks *et al.* [2]. There, a WP detector, in the form of a current carrying narrow conducting constriction (a Quantum Point Contact, QPC), was placed in close proximity to one of the two paths of the electron interferometer [3]. Electrons passing in this path affected the current of the detector, and even though the change in current was not actually measured, the mere possibility to measure it and obtain WP information was sufficient to partly dephase the interferometer. Contrary to that, in our present experiment, only a quantum mechanical phase is being added to the electrons that pass a QPC detector - **without affecting** the current - hence, WP information cannot be readily extracted. In spite of this 'difficulty' strong dephasing of the nearby interferometer is observed. Contrary to common intuition, we show that dephasing the detector does not affect its detection properties.

Before constructing the actual setup of the experiment we consider its basic requirements. Interference strength in the interferometer can be expressed in terms of a 'dephasing rate' induced by the detector times an 'effective time of interaction' between the interferometer and the detector. There are two equivalent ways to estimate the dephasing rate. The first is via studying the effect of the electron in the interferometer on the state of the detector, and the second is that of the detector on the interferometer. Here we emphasize first the former, easier to estimate, approach, based on a theory developed by Stern *et al*. [4] and applied to a QPC detector in Ref. 2. This approach relates the dephasing rate to the overlap of the two possible detector states: $\langle \chi_w | \chi_{wo} \rangle$, where the states $|\chi_w\rangle$ and $|\chi_{wo}\rangle$ correspond to the state of the detector with and without an electron in the nearby path, respectively. A small overlap indicates nearly orthogonal detector states and clear WP determination, hence strong dephasing.

The overlap of detector states, with current flowing through the QPC detector, can be expressed in terms of the single electron states in the QPC and raised to the power $N$, with $N$ the number of impinging electrons on the QPC during the 'effective time of interaction'. Each single electron state can be expressed as $t_d|t\rangle + r_d|r\rangle$, where $t_d$ and $r_d$ are the complex transmission and reflection amplitudes, and $|t\rangle$ and $|r\rangle$ are the quantum states of the transmitted and reflected electrons, respectively. For weak interaction between detector and interferometer the dephasing rate, $1/\tau_\varphi$, can be shown to have the form [2,5]:

$$\frac{1}{\tau_\varphi} = \frac{eV_d}{8\pi h}\frac{(\Delta T_d)^2}{T_d(1-T_d)} + \frac{2eV_d}{h}T_d(1-T_d)\sin^2(\gamma/2) \quad , \tag{1}$$

where $T_d=|t_d|^2$ is the transmission probability through the QPC, $V_d$ is the voltage applied to the QPC, $\Delta T_d$ is the change in the transmission coefficient, and $\gamma = \Delta\theta_t - \Delta\theta_r$ is the induced phase change, both are due to an electron dwelling in the nearby path of the interferometer. The phases $\Delta\theta_t$ and $\Delta\theta_r$ are the respective phase changes in the transmitted and reflected waves. Since in our experiment $\Delta T_d=0$ we address only the second term in Eq. (1). While the term $\Delta T_d$ is determined directly via conductance measurement of the QPC the phase change, $\gamma$, cannot be measured within the QPC detector. Such phase change can be measured, in principle, via an interference experiment, as the second term in Eq. (1), being periodic in $\gamma$, suggests. Interference, taking place between the transmitted wave with magnitude $\sqrt{T_d}$ and the reflected wave with magnitude $\sqrt{1-T_d}$, leads to a change in the conductance of the appropriately constructed interference loop. However, our detector does not contain such an interference loop and cannot provide such information. Moreover, this interpretation implies that coherency of the detector is required! We address this issue later. A second way to understand the dephasing process is to inspect the effect of the detector on the interferometer. This was done by Levinson [6] who calculated the inelastic scattering rate due to the non-equilibrium charge fluctuations in the near by detector. These fluctuations, being proportional to $T_d(1-T_d)$, result from partitioning of the current by the QPC and are the ubiquitous *shot noise* resulting from the *granular* nature of the electrons [7]. Equation (1) is still puzzling. While the second term in Eq. (1) is in agreement with the above understanding, the first term is counterintuitive. Being inversely proportional to the shot noise, we may interpret it as the limiting parameter that prevents accurate WP determination by the detector.

The realization of the experiment proceeds as follows. The condition $\Delta T_d=0$, $\gamma=0$ can be easily realized with electrons in a high magnetic field (being in the so-called: Quantum Hall Effect regime). Under these conditions most current flows in a chiral motion along trajectories that follow the edges of the sample (called edge states). A QPC constriction, which effectively brings the two edges of the sample close together, enables scattering between the edges, and serving as an electron 'beam splitter' (Fig. 1). The nearby interferometer induces a phase change in one of the propagating edge states (in the transmitted wave) hopefully yielding the desired WP information.

A standard two-path interferometer, employed in Ref. 2, is ineffective in the presence of a high magnetic field since it prevents electrons from choosing either path with nearly equal probability. Hence, we adopted an electronic version of the optical Fabry-Perot interferometer. A relatively simple version is the *quantum dot* (QD) [3,8], which is a small electronic trap connected with two leads to electron reservoirs, capturing electrons for a relatively long dwell time as they bounce back and forth inside the potential well that forms the trap. The interference, among the many bouncing paths, leads to sharp resonances in the transmission through the QD (hence in the conductance) as a function of energy. Experimentally, the obtained peaks in the conductance (known as Coulomb Blockade (CB) peaks) are thermally broadened due to the Fermi distribution in the leads with width considerably larger than the intrinsic resonance width $\Gamma_i$. We used two QDs in series (*double quantum dot* system (DQD)), where degeneracy between the resonant levels in both dots is necessary for conduction, hence enabling CB peaks $2\Gamma_i$ wide [9]).

When an electron enters and dwells in the DQD interferometer it charges the DQD and causes a deflection of the nearby edge-state away from the DQD (see Fig. 1). The deflected path adds an extra phase to the transmitted electrons, $\Delta\theta_t$, without affecting the transmission probability of the QPC ($\Delta T_d$=0). It is not difficult to understand that longer dwell time trajectories in the DQD will be affected more by the detector than those with shorter dwell times, making them *distinguishable* and leading to dephasing of the DQD. We expect then to measure a reduction of the peak height and a broadening of the peak width to $\Gamma_i + \hbar/\tau_\varphi$. A small shift in the energy of the peak is also expected but since there are other causes for a shift we do not make use of it. Note that in our experiment we can control the detector's transmission $T_d$, the voltage applied to the detector $V_d$, and the magnetic field (that affects $\gamma$), hence controlling the induced dephasing rate.

The structure, fabricated in a high mobility two dimensional electron gas, is schematically shown in Fig. 1. The DQD is being tuned with its gates to form resonance peaks 'above' the two-dimensional plane spanned by the two plunger gate voltages, $V_{p1}$ and $V_{p2}$. The resonance peaks are located on a hexagonal lattice (Fig. 2a); a well-known fingerprint of the DQD in the CB regime [10]. A single, magnified peak, with its contour at half maximum, is shown in Fig. 2b. We use the area in this contour as a measure of the dephasing rate.

Dephasing is being studied, at low temperatures, as a function of $T_d$ and $V_d$. A high magnetic field is applied leading to one or two filled Landau levels, namely, a *filling factor* one or two (FF = 1, 2). We present here results for FF=2 but note, however, that at the relatively high $V_d$ being used, the two Landau levels cannot be resolved. Figure 3a shows the dependence of one CB 'contour area' on $T_d$ for a fixed $V_d$. The measured area qualitatively follows the expression $T_d(1-T_d)$ in Eq. (1). Similarly, the peak height has an inverse dependence on that expression (Fig. 3b). A nearly linear dependence of the dephasing rate on the applied voltage $V_d$, at $T_d$=0.7, is shown in the inset of the Fig. 3. Similar results were obtained from the behavior of other peaks, in a few different devices, and for FF=1. Note that our discussion must remain qualitative since we have no theory to express the 'contour area' or the height of the CB peaks of the DQD as a function of the induced dephasing rate by the detector. The results in Fig. 3 clearly show that indeed phase change in the detector leads to dephasing of the interferometer even though an interference experiment was not performed in the detector.

Since according to our initial understanding WP detection relies on interference between transmitted and reflected waves, an important question naturally rises: Must the detector be phase coherent in order to dephase the interferometer? At first glance it seems that an incoherent detector will prevent WP detection. However, as it turns out this is not the case. This can be shown by adding an artificial *dephasor* in the path of the transmitted wave before it reaches the DQD, thus coupling the detector the *dephasor's* degrees of freedom. The single particle states of the 'modified detector' (detector+*dephasor*) can now be written in the form: $t_d|t\rangle\,|\pi_t\rangle + r_d|r\rangle\,|\pi_r\rangle$, where $|\pi_t\rangle$ and $|\pi_r\rangle$ are the states of the *dephasor* for an electron being transmitted or reflected by the QPC, respectively. It can then be shown that the overlap factor $\langle\chi_w|\chi_{wo}\rangle$ and consequently the dephasing rate remain the same as without the *dephasor*. In spite the fact that the presence of such *dephasor* prevents obtaining WP information by interfering $|t\rangle$ and $|r\rangle$, still, dephasing in the interferometer takes place! It might be easier to grasp this counter intuitive fact if we realize that an ideal *dephasor*, substituted in the current path, is not expected to affect the magnitude of charge fluctuations in the detector (due to current conservation).

In order to verify the role of detector coherency we introduced a floating Ohmic contact between the QPC and the DQD [11] (Fig. 4a). The Ohmic contact serves as a thermal bath, emitting the edge states that enter it totally dephased. A biased gate in front of the Ohmic contact allows removing it from the path of the transmitted edge state. Figure 4b shows the dependence of the peak 'contour area' on $T_d$, for an Ohmic contact in and out the path of the edge state. Surprisingly, we find an unexpected decrease in the dephasing rate in the presence of the Ohmic contact. We attribute this decrease to the finite capacitance of the Ohmic contact (which is difficult to minimize), 'shorting to ground' high frequency components of the shot noise, preventing them from arriving at the DQD. Indeed, adding an external capacitor decreased the dephasing rate even more. While charge fluctuations in the edge channel that leave the QPC have all frequency components, $f<10^{12}$ Hz, it is estimated that after leaving the Ohmic contact only frequencies below some $10^7$ Hz remain. Even though this is a negligible part of the spectrum it is reasonable to believe that the lowest part of the spectrum leads to phase smearing among the stream of the electrons passing the DQD. In other words, each electron, dwelling in the DQD for only a few nanoseconds, emerges with a random phase. With these assumption in mind we believe that the results with and without the artificial *dephasor* are consistent with Eq. (1).

The results presented here show that in an interferometer-detector experiment (*which path* experiment), even in the presence of strong induced dephasing, there are times when WP information cannot be readily extracted from the detector. In such cases it might be easier to study the direct effect of the detector on the interferometer. Such experiments can be exploited to provide new information on both the *dephasor* and the *dephasee*. For example, having the detector in the Fractional QHE regime, where current carrying charges are fractionally charged, can shed light on the interaction between fractional charges and the mundane electrons in the interferometer, thus leading to a new type of entangled states.

We thank D. Mahalu for her extremely valuable help in processing the devices, and to J. Yang and H. Moritz for their help during all the experiments. We are in debt to S. Gurvitz and L. Stodolsky who introduced the problem to us, and to Y. Aharonov who opened our eyes to see what is so obvious to him. We benefited from many discussion with J. Imry, Y. Levinson, S. Levit and A. Stern. On of us (E.B.) thanks to M. Buttiker and Y. Meir for useful discussions. The work was partly supported by a MINERVA grant.

**Reference**

[1] Bohr, N., Discussion with Einstein on Epistemological Problems in Atomic Physics, in *Albert Einstein: Philosopher - Scientist* (Ed. P. A. Schilpp) 200-241 (Library of Living Philosophers, Evanston, 1949).

[2] E. Buks, R. Schuster, M. Heiblum, D. Mahalu and V. Umansky, "Dephasing in electron Interference by *Which Path* Detector", Nature **391**, 871 (1998).

[3] Schuster, R. Buks, E., Heiblum, M., Mahalu, D., Umansky, V., & Shtrikman, Hadas, "Phase Measurement in a Quantum Dot via a Double-Slit Interference Experiment", Nature **385**, 417-420 (1997).

[4] Stern, A., Aharonov, Y., & Imry, Y., "Phase Uncertainty and Loss of Interference: A General Picture", Phys. Rev. A **41**, 3436-3448 (1990).

[5] Some version of Eq. (1) appears in: Stodolsky, L., "Measurement Process in a Two-Barrier System", quant-ph/9805081; Buks, E., "Interference and Dephasing in Mesoscopic Systems", PhD thesis; Gurvitz, S. A., "Measurements with a Noninvasive Detector and Dephasing Mechanism", Phys. Rev. B, in press, cond - mat/9706074; Aleiner, I. L., Wingreen, N. S., & Meir, Y., "Dephasing and the Orthogonality Catastrophe in Tunneling through a Quantum Dot: the "Which Path?" Interferometer", Phys. Rev. Lett. **79**, 3740 (1997).

[6] Levinson, Y., "Dephasing in a Quantum Dot due to Coupling with a Quantum Point Contact", Europhys. Lett. **39**, 299 (1997); Levinson, Y., to be published.

[7] A review on shot noise in mesoscopic systems: Reznikov, M., de Picciotto, R., Heiblum, M., Glattli, D. C., Kumar, A., & Saminadayar, L., "Quantum Shot Noise", Superlattices and Microstructures **23**, 901-915 (1998).

[8] Van Houten, H., Beenakker, C. W. J., & Staring, A. A. W., Coulomb Blockade Oscillations in Semiconductor Nanostructures, in *Single Charge Tunneling - Coulomb Blockade Phenomena in Nanostructures*, (Eds. Grabert, H., & Devoret, M. H., Plenum Press, New York, 1992).

[9] Livermore, C., Crouch, C. H., Westervelt, R. M.,Campman, K. L., & Gossard, A. C., "The Coulomb Blockade in Coupled Quantum Dots", Science **274**, 1332-1335 (1996).

[10] van der Vaart, N. C., Godijn, S. F., Nazarov, Y. V., Harmans, C. J. P. M., & Mooij, J. E., "Resonant-Tunneling through 2 Discrete Energy-States", Phys. Rev. Lett. **74**, 4702-4705 (1995).

[11] Buttiker, M., Andrew, M. M., "Charge Relaxation and Dephasing in Coulomb Coupled Conductors", cond-mat/9902320

## Figure Captions

**Figure 1**: A schematic of the DQD interferometer and a QPC detector. The device is fabricated in a high mobility two dimensional electron gas embedded in a GaAs-AlGaAs heterostructure with a low temperature (at ≈1 K) mobility of the electrons $5x10^5$ $cm^2/V$-sec. The QPC, the DQD, and the trajectories of the edge states were defined by depositing metal gates on the surface of the structure and biasing them negatively. This leads to depletion of electrons underneath the gates, forming impenetrable potential barriers, serving as potential and guiding walls. In the presence of a strong perpendicular magnetic field (5-10 Tesla) and applied voltage $V_d$ , electrons, emanating from an Ohmic contact (on the left), travel in edge states toward the QPC. The magnetic field is tuned so that one or two edge states coexist near the boundary (the so-called filling factor one or two, respectively). The edge states are partly transmitted ($|t\rangle$) and partly reflected ($|r\rangle$) by the QPC. The transmitted edge state is Coulombically coupled to the near by DQD interferometer, which is in turn connected to its own circuit (with source *S* and drain *D*). The DQD, being weakly coupled to its own leads (via two, slightly pinched off, QPCs), is tuned to resonance via voltages to the two *plunger* gates, $V_{p1}$ and $V_{p2}$, each tunes the resonance in one QD. Whenever an extra electron is added to the DQD the nearby edge state is slightly diverted (dashed line) thus acquiring additional phase, $\gamma$. This phase change distinguishes between the different possible trajectories in the DQD, each with its own dwell time.

**Figure 2**: (a) Conductance of the DQD interferometer as a function of the two plunger gate voltages $V_{p1}$ and $V_{p2}$. The bright spots correspond to CB peaks. A resonant peak

occurs in the degeneracy point of three charge states allowing transport of one electron from one side of the DQD to the other: n electrons in the left dot and m electrons in the right dot; n+1 electrons in the left dot and m electrons in the right dot; n electrons in the left dot and m+1 electrons in the right dot; and back to n and m electrons in each dot. The hexagonal ordering is a known property of a DQD system [9]. (b) Magnified view of one CB peak. The dashed line is a contour drawn at half maximum of the peak height. Due to the asymmetry of the peak shape we use the area enclosed by this contour as a measure of the peak width.

**Figure 3**: (a) The area of the contour at half peak height as a function of the transmission probability, $T_d$, for two values of applied bias $V_d$ = 0 and 2 mV. The dependence qualitatively agrees with the expected $T_d(1-T_d)$. The measurement is done at a filling factor 2, namely, with two spin split levels (two edge states). Changing the bias on the gates that form the QPC varies the conductance and the transmission coefficient of the QPC. Note that the large applied $V_d$ does not allow resolving the splitting of the two edge states, hence we use the expression for conductance $T_d \cdot 2e^2/h$ to obtain $T_d$. Inset: The dependence of the contour area on the applied voltage, $V_d$, for $T_d$ = 0.7. As expected this dependence is nearly linear. (b) The dependence of the peak height (in units of conductance) on the transmission probability, $T_d$, for two values $V_d$ = 0 and 2 mV. This dependence also agrees with the presented model. Note, that in addition to the change in the peak height there is also a change in its position, which is not shown.

**Figure 4**: (a) A schematic of the experimental setup with a floating Ohmic contact introduced in the transmitted wave. The Ohmic contact serves as a dephasor for the transmitted electrons. A gate in front of the Ohmic contact allows removing this contact from the electrons path. (b) The area of the contour at half peak height as a function of the transmission probability, $T_d$, with and without the Ohmic contact in the electrons path. The decrease in the dephasing rate is attributed to the non-negligible capacitance of the Ohmic contact that shorts the high frequency components of the shot noise to ground, preventing them from participating in the dephasing of the DQD.

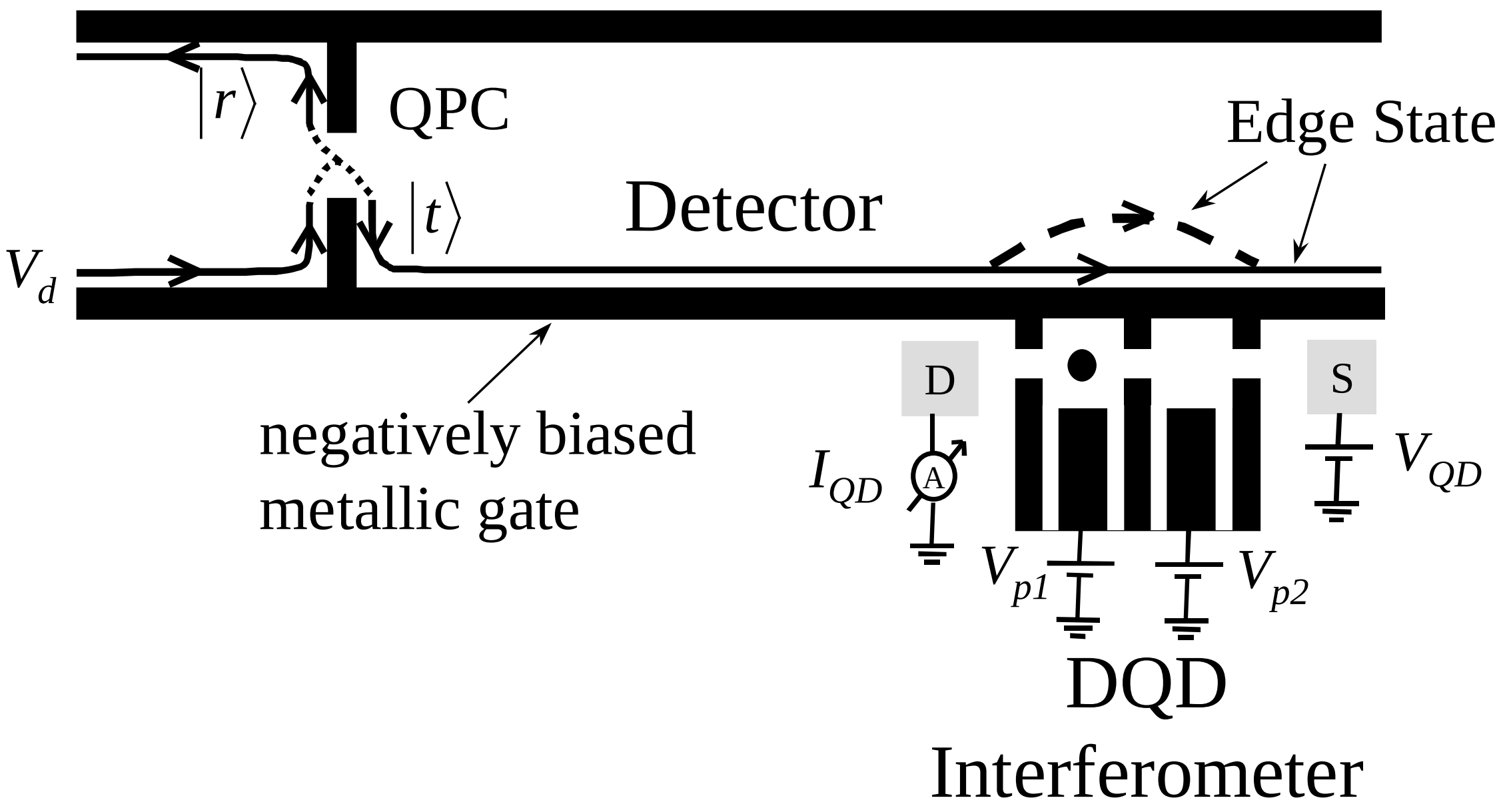

Figure 1

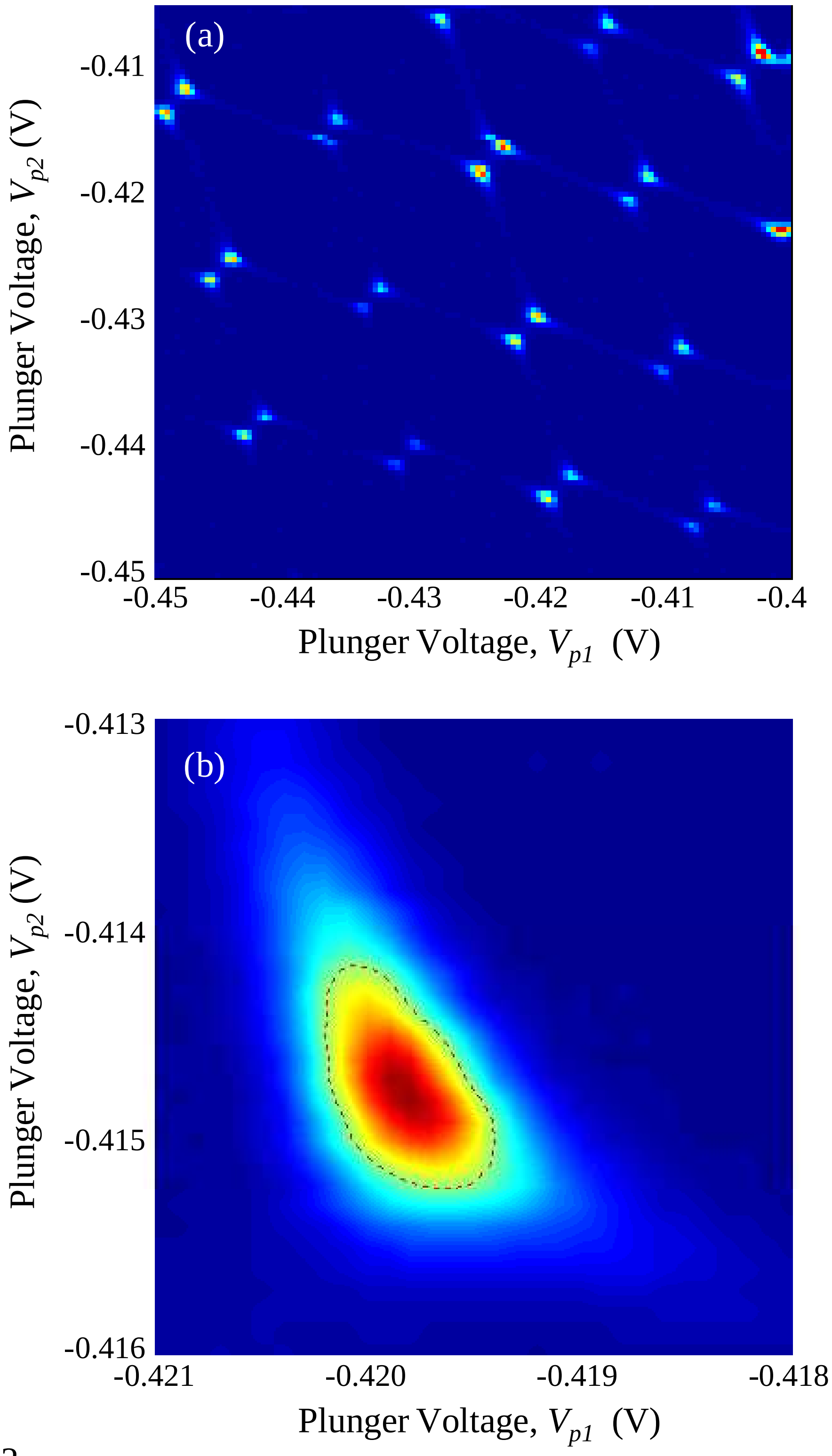

Figure 2

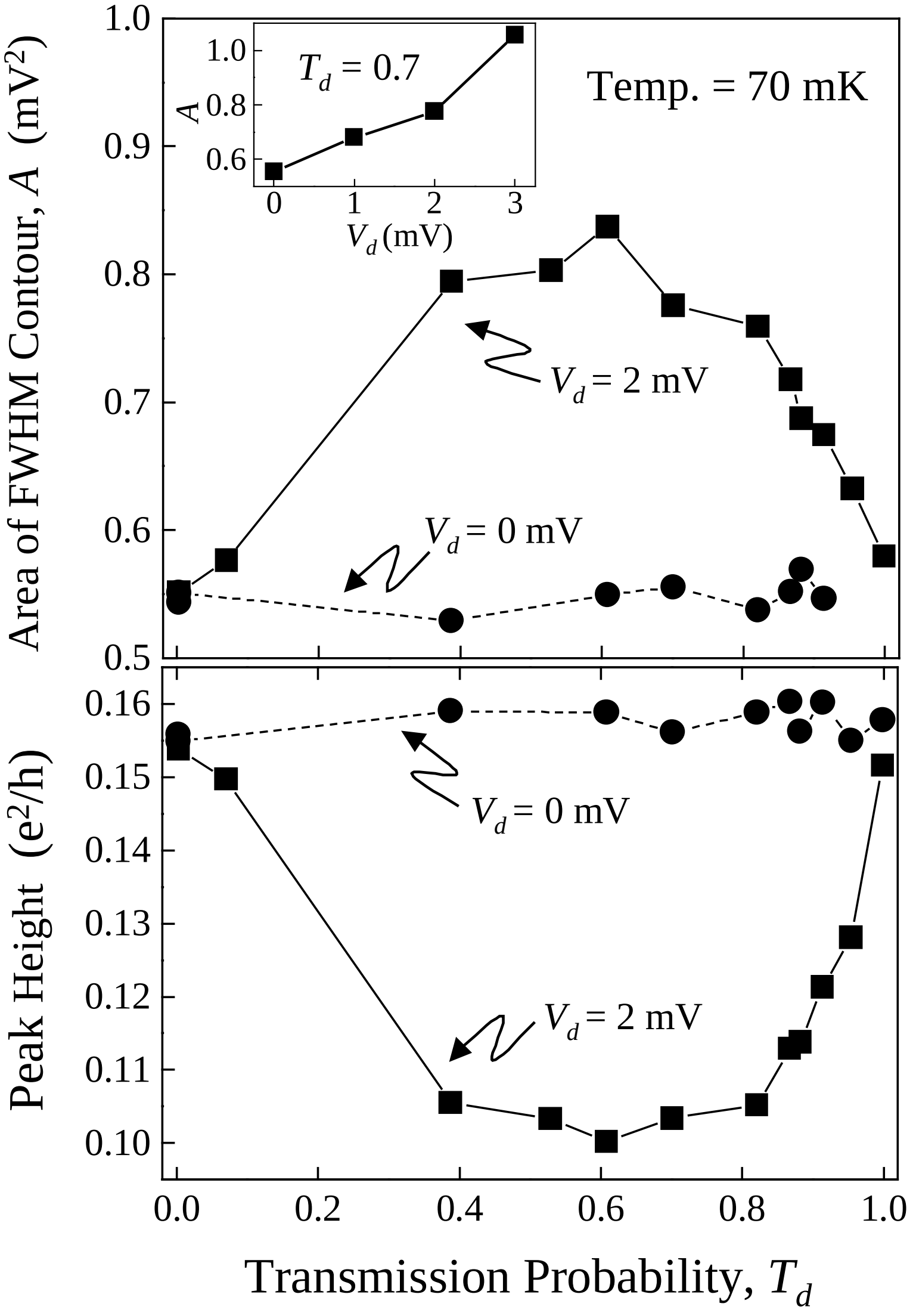

Temp. = 70 mK
$T_d$ = 0.7
A
$V_d$ (mV)
$V_d$ = 2 mV
$V_d$ = 0 mV
Area of FWHM Contour, $A$ (mV$^2$)
Peak Height (e$^2$/h)
$V_d$ = 0 mV
$V_d$ = 2 mV
Transmission Probability, $T_d$

Figure 3

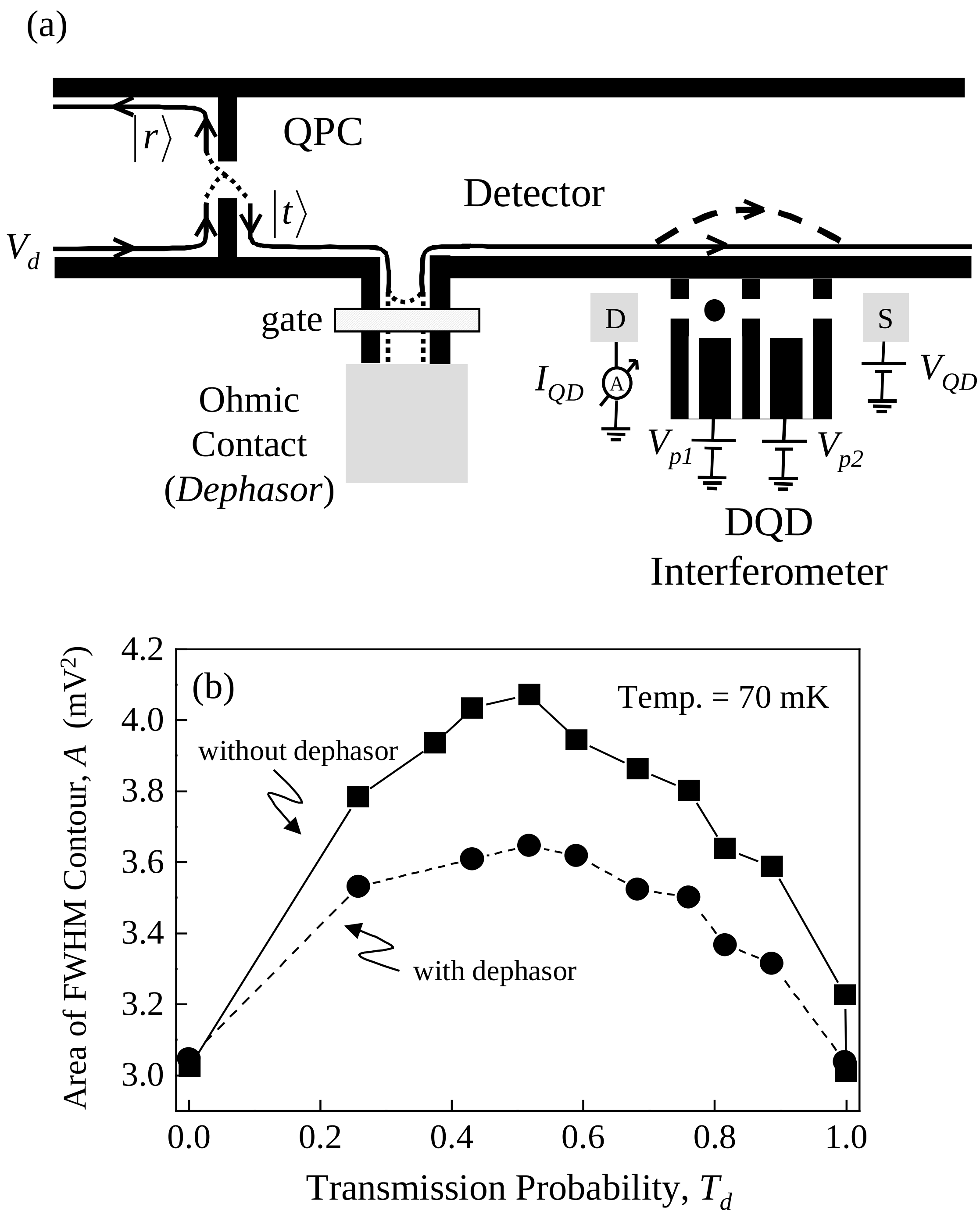

Figure 4